\documentclass[a4paper,11pt]{article}
\usepackage[utf8]{inputenc}
\pdfoutput=1 
\usepackage{jcappub} 
\usepackage{afterpage}
\allowdisplaybreaks
\usepackage[T1]{fontenc}
\usepackage{subcaption}
\usepackage{siunitx}
\DeclareSIUnit\parsec{pc}
\usepackage[nameinlink]{cleveref}
\crefname{figure}{figure}{figures}

\definecolor{lightgray}{gray}{0.9}
\usepackage{listings}
\newcommand{\av}[2][]{\left\langle #2\right\rangle_{#1}}

\newcommand\mL{\mathcal{L}}
\newcommand\mP{\mathcal{P}}
\newcommand\mZ{\mathcal{Z}}
\newcommand\mD{\mathcal{D}}
\newcommand\mS{\mathcal{S}}

\title{A cosmic glitch in gravity}

\author[a,b]{Robin Y. Wen,}
\author[c]{Lukas T. Hergt,}
\author[d,e,f]{Niayesh Afshordi,}
\author[c]{Douglas Scott}

\affiliation[a]{Department of Applied Mathematics, University of Waterloo, \\Waterloo, ON N2L 3G1, Canada}

\affiliation[b]{California Institute of Technology, Pasadena, CA 91125, USA}

\affiliation[c]{Department of Physics \& Astronomy, University of British Columbia, Vancouver, BC V6T 1Z1, Canada}

\affiliation[d]{Waterloo Centre for Astrophysics, University of Waterloo, Waterloo, ON, N2L 3G1, Canada}

\affiliation[e]{Department of Physics \& Astronomy, University of Waterloo, Waterloo, ON N2L 3G1, Canada}

\affiliation[f]{Perimeter Institute For Theoretical Physics, Waterloo, ON N2L 2Y5, Canada}

\emailAdd{y52wen@uwaterloo.ca}
\emailAdd{lthergt@phas.ubc.ca}
\emailAdd{nafshordi@pitp.ca}
\emailAdd{dscott@phas.ubc.ca}

\abstract{We investigate a model that modifies general relativity on cosmological scales, specifically by having a `glitch' in the gravitational constant between the cosmological (super-horizon) and Newtonian (sub-horizon) regimes, as motivated e.g.\ in the Ho\v{r}ava--Lifshitz proposal or in the Einstein-aether framework.  This gives a single-parameter extension to the standard $\Lambda$CDM model, which is equivalent to adding a dark energy component, but where the energy density of this component can have either sign.  Fitting to data from the \textit{Planck} satellite, we find that negative contributions are, in fact, preferred.  Additionally, we find that roughly one percent weaker superhorizon gravity can somewhat ease the Hubble and clustering tensions in a range of cosmological observations, although at the expense of spoiling fits to the baryonic acoustic oscillation scale in galaxy surveys. Therefore, the extra parametric freedom offered by our model deserves further exploration, and we discuss how future observations may elucidate this potential \textit{cosmic glitch in gravity}, through a four-fold reduction in statistical uncertainties.}

\begin{document}
\maketitle
\flushbottom

\section{Introduction}

Albert Einstein's proposal of the general theory of relativity~(GR) that replaced the Newtonian gravitational force with a dynamical living, breathing spacetime geometry has become the cornerstone of modern physics over the past century, having passed all the empirical tests thrown at it --- too many to count here --- across diverse scales and regimes~\cite{will2020einstein}. Furthermore, a range of mathematical theorems prove the uniqueness of general relativity, subject to conditions, namely covariance (or absence of a preferred frame of reference) and minimal number of degrees of freedom (or only two polarizations for gravitational waves)~\cite{1972JMP....13..874L,1974JMP....15..708K,1976AnPhy..96...88H,khoury2012spatially,krasnov2021pure}. Nevertheless, several motivations, most notably the inconsistency of the postulates of quantum theory with those of general relativity, have pushed theorists to explore beyond this minimal set of assumptions. Furthermore, the existence of a preferred cosmological reference frame (where the cosmic microwave background has no dipole), which in the standard picture carries no physical meaning, suggests scrutinising alternatives in which there is a genuinely special cosmological frame.  A consequence of such an idea would be that minimal deviations from general relativity would be expected only on cosmological scales~\cite{07CuscutonCosmo,mukohyama2019minimally}, or else would require new degrees of freedom (as in modifications that might occur in the strong gravity regime near black holes). 

In lieu of introducing a new length scale, some such theories predict a `glitch' between the gravitational constant that governs cosmology on super-horizon scales $G_\mathrm{cosmo}$ (through the Friedmann equation) and Newton's constant of gravitation, $G_\mathrm{N}$ that governs the inverse square law on sub-horizon scales. For example, in the Ho\v{r}ava--Lifshitz proposal for a Lorentz-violating theory of gravity~\cite{hovrava2009quantum}, the low energy non-projectible theory predicts ${G_\mathrm{N}/G_\mathrm{cosmo} = 1-\frac{3}{2}(\lambda-1)}$~\cite{09CuscutonHovrava}, where $\lambda$ is the coefficient of the mean extrinsic curvature squared term in the theory (with $\lambda =1$ in GR). Alternatively, within the Einstein-Aether framework, the parameter $\lambda-1$ is replaced by $c_2$ which quantifies the coupling of geometry with the divergence of the aether flow~\cite{jacobson2010extended}. Indeed, it has been shown that the low energy non-projectible Ho\v{r}ava--Lifshitz proposal and the Einstein-aether theory with only $c_2$ non-vanishing are physically identical to the quadratic cuscuton theory~\cite{07CuscutonCosmo}, an incompressible scalar field theory with a quadratic potential~\cite{09CuscutonHovrava}. Consequently, it can be formally proven that none of these theories is distinguishable from general relativity in asymptotically flat spacetimes~\cite{Loll:2014xja}, and thus can only be tested on scales comparable to (or larger than) the cosmological horizon~\cite{Robbers:2007ca}. Using the Friedmann equation
\begin{equation}
H^2= \frac{8\pi G_\mathrm{cosmo}}{3}\rho_\mathrm{tot} 
= \frac{8\pi G_\mathrm{N}}{3}\left[\rho_\mathrm{tot}+\left(1-\frac{G_\mathrm{N}}{G_\mathrm{cosmo}}\right)\rho_\mathrm{crit}\right], 
\end{equation}
it is possible to reinterpret this cosmic glitch in gravity as a dark energy component with a constant density relative to critical density:
\begin{align}
 \Omega_\mathrm{g} = \frac{\rho_{\rm DE} -\rho_\Lambda}{\rho_\mathrm{crit}} =  1 -\frac{G_\mathrm{N}}{G_\mathrm{cosmo}}  \label{eq:G_cosmo}.
\end{align}

Over the past couple of decades, experimental results have established cold dark matter with a cosmological constant~$\Lambda$ as the standard model of cosmology (referred to as $\Lambda$CDM).
Although the $\Lambda$CDM model has been extremely successful in explaining a wide range of cosmological observables with only a small number of free parameters, there are some observations that might point to deficiencies in the model. One recent example is an apparent discrepancy between the measurement of $^4$He abundances in ten extremely metal poor galaxies by the EMPRESS collaboration~\cite{matsumoto2022empress}. Here, standard Big Bang predictions can be reconciled if $\Omega_\mathrm{g} = -0.085\pm0.027$ (1$\,\sigma$) during nucleosynthesis~\cite{22EMPRESSVIIICuscuton}. But is this the only consequence?  The primary goal of this paper is to examine the case for a non-vanishing cosmic glitch $\Omega_\mathrm{g}$ for other cosmological observables. 

Despite the tremendous success of $\Lambda$CDM, the increasing precision of cosmological measurements has revealed several discrepancies between different cosmological probes~\cite{22CosmoIntertwined}, with the most prominent being the different estimates of the Hubble parameter~$H_0$. This Hubble tension refers to the difference between the estimates of $H_0$ based on the distance ladder and the determinations inferred from cosmic microwave background (CMB) anisotropies interpreted through a cosmological model.   This discrepancy has now reached perhaps $5\,\sigma$ significance between the value derived from \textit{Planck} satellite data~\cite{18Plancklikelihood,18Planckparameter,Tristram2023} under the $\Lambda$CDM model and the value from the SH0ES project using Cepheid-calibrated Type Ia supernovae~\cite{2019SH0ES,21SH0ES,22SH0ES}. 

In light of this tension, growing efforts have been made to explore the possibility of new physics beyond the $\Lambda$CDM model. Many models of non-minimal dark energy beyond a cosmological constant have been proposed and developed, with the goal of solving the Hubble tension in addition to other phenomenological and theoretical motivations. A popular subclass of these models is the early dark energy~(EDE) idea, where the modified dark energy component has the greatest impact during the early phase of the Universe, before cosmological recombination~\citep{Poulin2023}. Several EDE models have been developed and tested over the years~\cite{06DoranEDE,13PettorinoEDE,15PlanckDEMG,19PoulinEDE,20Ye_EDEAds}, with various degrees of success on addressing the Hubble tension, depending on the particular model and the data used \cite{19PoulinEDE,20Ye_EDEAds,22ACTPreRecombEDE,20HillEDE,21Jiang_AdSEDE}.

An additional tension seen in cosmological observables is the inferred value of the parameter $S_8\equiv\sigma_8(\Omega_\mathrm{m}/0.3)^{0.5}$ between galaxy weak lensing and CMB measurements~\citep[e.g.][]{22Abdalla_tension}.  More generally, there is a tension between galaxy clustering measurements and CMB measurements in a 2-dimensional plane~\citep[e.g.][]{19Handley}. EDE models that try to address the $H_0$ tension typically do not improve this clustering tension. We will investigate these two tensions for our CGG model. 

The CGG parameterisation that we adopt here is similar to the particular EDE model proposed by Doran and Robbers~\cite{06DoranEDE}, which has been constrained in Ref.~\cite{15PlanckDEMG} with \textit{Planck} 2015 data and further extended with more parameters in Refs.~\cite{13PettorinoEDE,21Gomez-ValentEDEplateau}. In these previous works, the density of the EDE component is always assumed to be positive, which leads to a particularly tight (though one-sided) constraint on $\Omega_\mathrm{g}$~\cite{15PlanckDEMG}. In contrast, we shall relax the positive requirement on $\Omega_\mathrm{g}$ and allow it to take negative values in order to fully explore the parameter space and exploit the degeneracy between $H_0$ and $\Omega_\mathrm{g}$. We constrain the model using the \textit{Planck} 2018 data~\cite{18Planckparameter} and discuss the prospect of alleviating the $H_0$ and $S_8$ tensions. 

The remainder of the paper is organised as follows. In \cref{sec:model}, we present our phenomenological implementation of the CGG model and its effects on the CMB power spectra. We then provide the constraints on $\Omega_\mathrm{g}$ using the cosmological data in \cref{sec:posterior-results}, including a discussion of the Hubble constant (\cref{sec:H0tension}) and clustering (\cref{sec:S8tension}) tensions, as well as consideration of Bayesian model comparisons between $\Lambda$CDM and CGG models (\cref{sec:model-comparison}). In \cref{sec:PR4} we discuss the impact of replacing the Planck18 likelihood, which comes from the Planck Public Release~3~(PR3), with the recent \texttt{LoLLiPoP} and \texttt{HiLLiPoP} likelihoods \cite{Tristram2023} from the Planck Public Release~4~(PR4). We provide forecasts on $\Omega_\mathrm{g}$ constraints with future cosmological measurements in \cref{sec:forecast}, discuss the possibility that the strength of the glitch might vary over time in \cref{sec:discussion}, and conclude in \cref{sec:conclusion}. Except where explicitly stated otherwise, all parameter uncertainties are given as $\pm1\,\sigma$, which corresponds to the 68$\,\%$ confidence interval for a Gaussian distribution.

\begin{figure}[tbp]
\centerline{\includegraphics[width=0.57\textwidth]{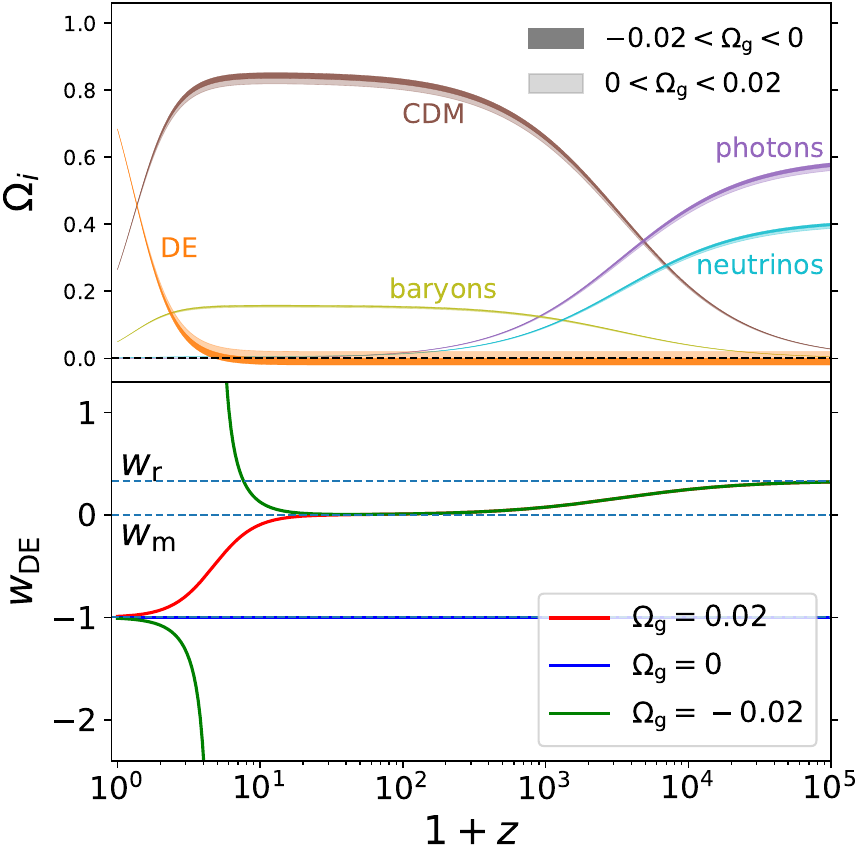}}
\caption{Evolution in the CGG model.  The upper panel shows the effects of $\Omega_\mathrm{g}$ on the energy density compositions $\Omega_i(z)$ for the photons, baryons, CDM, neutrinos (including both massive and massless species) and dark energy components, respectively. We use the dark (light) shaded region to indicate regions corresponding to negative (positive) $\Omega_{\rm g}$ values, relative to $\Lambda$CDM. The lower panel shows the equation of state for the dark energy component $w_\mathrm{DE}(z)$ for different values of $\Omega_{\rm g}$. The solid blue line tracks the evolution of a universe with $\Omega_\mathrm{g}=0$, corresponding to a standard $\Lambda$CDM cosmology with $w_\mathrm{DE}=w_\Lambda=-1$. The blue dashed lines show the equation of state for radiation ($w_\mathrm{r}=\frac{1}{3}$) and matter ($w_\mathrm{m}=0$), which our CGG component traces, respectively, in the different eras. In both panels, we use the best-fit parameters for the 2018 \textit{Planck} $TT,TE,EE$+low$E$+lensing likelihood given in Ref.~\cite{18Planckparameter}. For the CGG model, we choose $\Omega_\mathrm{g}=\pm 0.02$, which reflects the $\Omega_{\rm g}$ constraints obtained from the \textit{Planck} 2018 data (shown in Table~\ref{tab:Constraints} and Fig.~\ref{fig:Pl18CGG}).}
\label{fig:w-omega}
\end{figure}

\section{The phenomenological model}\label{sec:model}

In our model, following \cref{eq:G_cosmo}, the energy density of the {\it effective\/} dark energy component is given by
\begin{equation}
    \rho_\mathrm{DE}= \rho_{\Lambda}+\Omega_\mathrm{g}\rho_\mathrm{crit},\quad\text{where}\quad \rho_\mathrm{crit}=\frac{3H^2}{8\pi G_\mathrm{N}}\quad\text{and}\quad \rho_{\Lambda}=\frac{\Lambda}{8\pi G_\mathrm{N}}.
    \label{eq:density-CGG}
\end{equation}
In addition to the constant dark energy density $\rho_{\Lambda}$ in the $\Lambda$CDM model, we introduce an extra component that is proportional to the critical density of the Universe, with $\Omega_\mathrm{g}$ being the proportionality constant.\footnote{Here $\Omega_\mathrm{g}$ is a parameter that does not vary with time.} Our model, therefore, modifies the amount of gravity for the background. Since $\rho_\mathrm{crit}$ also includes $\rho_\mathrm{DE}$, we can isolate $\rho_\mathrm{DE}$ from \cref{eq:density-CGG} to obtain
\begin{equation}
    \rho_\mathrm{DE}= \frac{\Omega_\mathrm{g}\rho_\mathrm{nonDE}+\rho_{\Lambda}}{1-\Omega_\mathrm{g}},
    \label{eq:density-CGG-2}
\end{equation}
where $\rho_\mathrm{nonDE}=\sum_{i}\rho_{i}=\rho_\mathrm{m}+\rho_\mathrm{r}+\rho_\mathrm{n\nu}+\rho_\mathrm{m\nu}$, which represents the densities of matter, radiation and massless and massive neutrinos, respectively. We consider a flat cosmology with $\Omega_{K}=0$ throughout this paper, and we set the effective number of relativistic species\footnote{Note that for compatibility with older computation runs we are using the value $N_\mathrm{eff}=3.046$, here, even though the default value in \texttt{Cobaya} has been changed to $N_\mathrm{eff}=3.044$.} $N_\mathrm{eff}=3.046$ and we assume a single massive neutrino with a mass of $m_{\nu}=\SI{0.06}{\eV}$.
Note that our model described in \cref{eq:density-CGG} is entirely consistent with the inclusion of massless and massive neutrinos. 

Since $\rho_{\Lambda}$ dominates over $\rho_\mathrm{nonDE}$ in the late Universe, in \cref{eq:density-CGG-2} the dynamics of this modified DE model is similar to $\Lambda$CDM during late times, but with a slightly different dark energy density due to the denominator $1-\Omega_\mathrm{g}$, while the dynamics of $\rho_\mathrm{DE}$ is substantially altered in the early Universe by tracking the behaviour of radiation and matter densities, respectively, during the radiation-dominated and matter-dominated eras.

Using \cref{eq:density-CGG-2}, the equation of state parameter for our model can then be written as 
\begin{align}
    1+w_\mathrm{DE}=-\frac{1}{3}\frac{d\ln\rho_\mathrm{DE}}{d\ln a}=-\frac{\Omega_\mathrm{g}}{3(1-\Omega_\mathrm{g})\rho_\mathrm{DE}}\sum_{i}\frac{d\rho_{i}}{d\ln a}=\frac{\Omega_\mathrm{g}}{(1-\Omega_\mathrm{g})\rho_\mathrm{DE}}\sum_{i}\rho_i(1+w_i)\label{eq:w-CGG},
\end{align}
where $w_\mathrm{m}=0$, $w_\mathrm{r}=w_\mathrm{n\nu}=\frac{1}{3}$, and the equation of state for massive neutrinos can be computed numerically in the code \texttt{CAMB}~\cite{CAMB}. As illustrated in the lower panel of \cref{fig:w-omega}, the CGG model tracks the component that dominates the density of the Universe, with $w_\mathrm{DE}$ transiting from $\frac{1}{3}$ to $0$ and then close to $-1$ as the Universe evolves.  The transition of $w_\mathrm{DE}$ is smooth for positive $\Omega_\mathrm{g}$, and the dynamics of our $w_\mathrm{DE}$ is similar to the EDE parameterisation in Ref.~\cite{06DoranEDE}. However, for negative $\Omega_\mathrm{g}$ values, $\rho_\mathrm{DE}$ in \cref{eq:density-CGG-2} is negative at early times and becomes positive at late times (the orange region in the upper panel of \cref{fig:w-omega}). The transition of $\rho_{\rm DE}$ from negative to positive values causes the diverging behaviour of $w_\mathrm{DE}$ (the green line) in the lower panel of \cref{fig:w-omega}. Negative values of $\rho_\mathrm{DE}$ typically lead to ghost instabilities in dynamical dark energy models \cite{11QCD_ghost,15Fulvio_ghost}, but here we can consider the model with negative $\Omega_\mathrm{g}$ as an approximate phenomenological description of modified gravity models, as described above. Some other models with effective dark energy density switching from negative to positive have also been discussed in the literature \cite{20Ye_EDEAds,21Calderon_NegativeDE,20Akarsu_DE_signswitch}.

\begin{figure}[tbp]
    \centerline{\includegraphics[width=.57\textwidth]{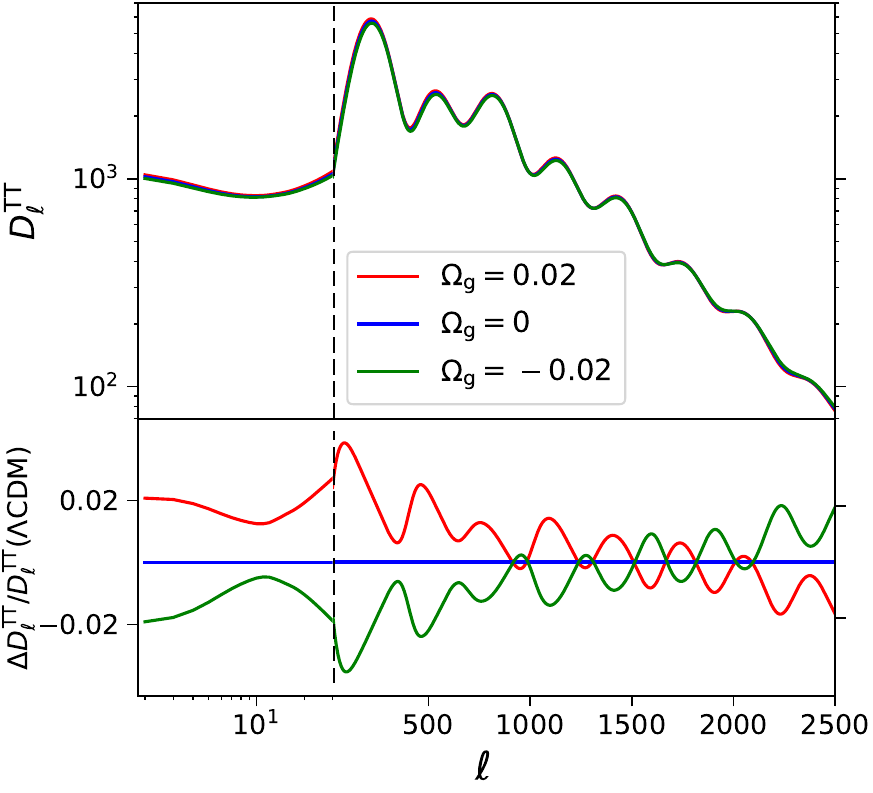}}
    \caption{Effects of $\Omega_\mathrm{g}$ on the CMB $TT$ power spectrum $D_{\ell}^{TT}=(\ell(\ell+1)/2\pi)C_{\ell}^{TT}$. We plot $D_{\ell}^{TT}$ using a logarithmic scale in the upper panel, and we show the relative difference between the $TT$ power spectrum of different models and that of $\Lambda$CDM ($\Omega_{\rm g}=0$). We use the best-fit parameters for the 2018 \textit{Planck} $TT,TE,EE$+low$E$+lensing likelihood given in Ref.~\cite{18Planckparameter}, along with $\Omega_\mathrm{g}=\pm0.02$, which reflects the $\Omega_{\rm g}$ constraints obtained from the \textit{Planck} 2018 data (shown in Table~\ref{tab:Constraints} and Fig.~\ref{fig:Pl18CGG}), for calculating the $TT$ power spectrum. We set the $x$-axis to have a logarithmic scale for low multipoles $\ell\leq 30$ and a linear scale for high multipoles $\ell>30$.}
     \label{fig:DlTT}
\end{figure}

There are motivations for adding a component that tracks the total density of the Universe, an example being the cuscuton model~\cite{07Cuscuton,07CuscutonCosmo,09CuscutonHovrava}, which naturally arises when trying to minimally modify GR by introducing a scalar field that does not propagate any new degrees of freedom. When the scalar field $\phi$ has a quadratic potential $V(\phi)=V_0+\frac{1}{2}m^2\phi^2$, the potential minimum $V_0$ contributes towards the cosmological
constant, while the quadratic term $\frac{1}{2}m^2\phi^2$ maintains a constant fraction of the total energy density of the Universe~\cite{07CuscutonCosmo}, which motivates the specific tracking behaviour introduced in \cref{eq:density-CGG} for $\Omega_\mathrm{g}>0$. For negative $\Omega_\mathrm{g}$ values we get $w<-1$ at late times, which suggests that the model crosses the so-called ``phantom~divide''. While a single (minimally-coupled) scalar field is gravitationally unstable at such a crossing, a double-field description has been applied to an extension of the \texttt{CAMB} framework to solve the perturbation equations~\cite{08HuPPFDE}, and we shall adopt this approach for our analysis.\footnote{The similarity of predictions from this framework, and the incompressible cuscuton limit was verified in Ref.~\cite{Robbers:2007ca}.} At the background level, a negative $\Omega_\mathrm{g}$ is equivalent to a smaller Newtonian gravitational constant in the Friedman equation, $G_\mathrm{cosmo} < G_\mathrm{N}$.

In order to obtain the CMB anisotropy and matter power spectra with non-zero $\Omega_\mathrm{g}$, we modify the \texttt{DarkEnergyFluid} module of \texttt{CAMB}~\cite{CAMB} where we numerically implemented \cref{eq:density-CGG-2,eq:w-CGG}. At the perturbation level, we consider the CGG component as a perfect fluid. Similarly to the approach taken in Refs.~\cite{22KrolewskiISWDE}, we make use of the parameterised post-Friedmann framework proposed in Ref.~\cite{08HuPPFDE,08FangPPFDE}, which is implemented as the \texttt{DarkEnergyPPF} module in \texttt{CAMB}, to allow $\Omega_\mathrm{g}$ to be negative. The parameterised post-Friedmann framework approximates the perturbation of the dark energy fluid assuming that it is smooth compared to dark matter, has vanishing anisotropic stress, and a rest frame speed of sound approximately equal to the speed of light~\cite{08FangPPFDE}, allowing the study of a generic dark energy sector with arbitrary $w(a)$. 

We plot the CMB temperature power spectrum~$D_{\ell}^{TT}$ with different $\Omega_\mathrm{g}$ values (while fixing all the other parameters to the \textit{Planck} best-fit values for $\Lambda$CDM) in \cref{fig:DlTT}. We see that a positive $\Omega_\mathrm{g}$ increases the integrated Sachs-Wolfe (ISW) effect at low multipoles $\ell$, along with a slight suppression of the small-scale power in $D_{\ell}^{TT}$, which is consistent with the predicted behaviour of the quadratic cuscuton~\cite{07CuscutonCosmo}. A negative $\Omega_\mathrm{g}$ reverses these trends.

\section{Constraints on the model} \label{sec:posterior-results}

\begin{figure}[tbp]
    \centerline{\includegraphics[width=\hsize]{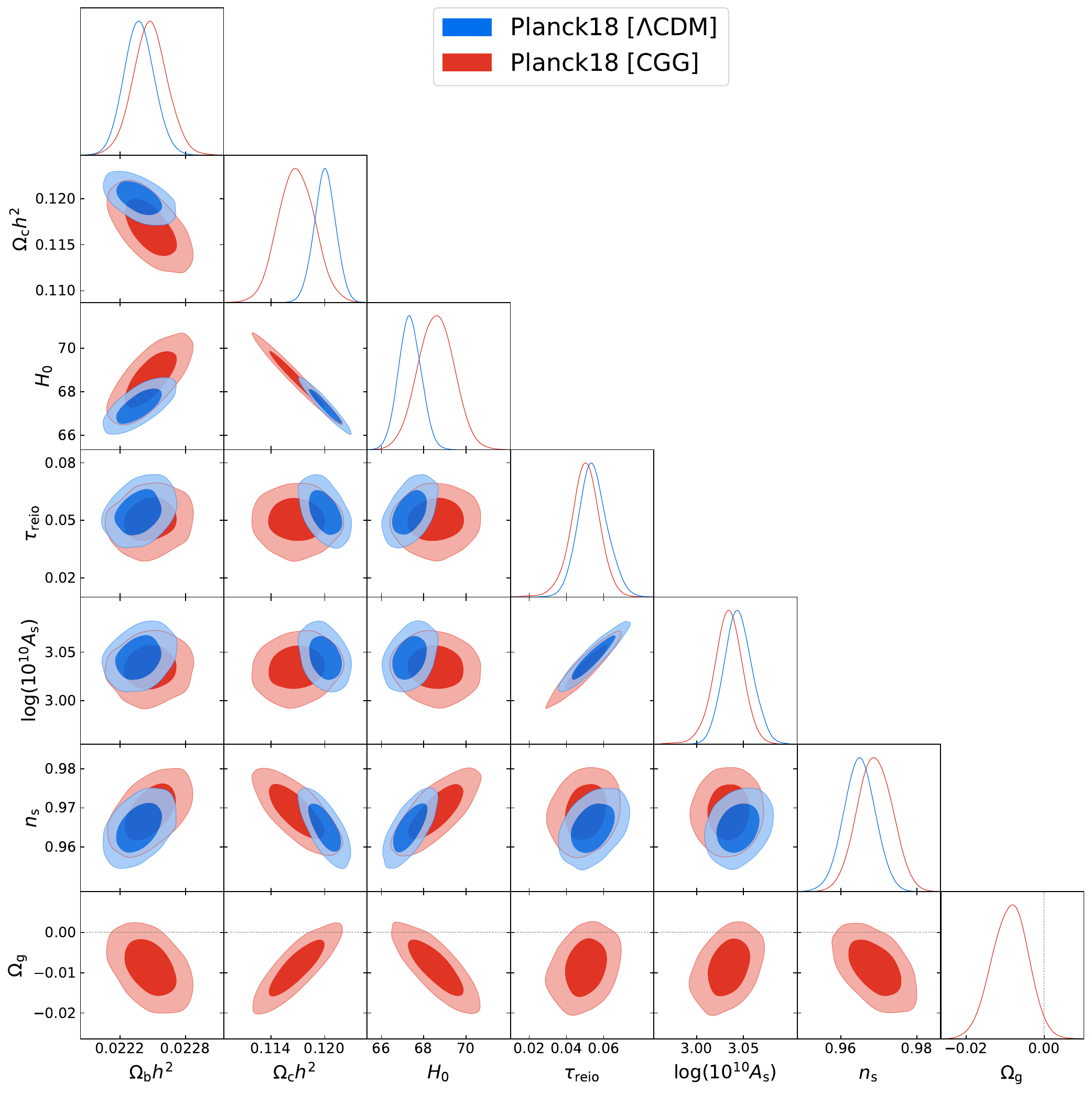}}
    \caption{Parameter constraints for the main seven cosmological parameters $\{\Omega_\mathrm{b}h^2,\Omega_\mathrm{c}h^2,H_0,\tau_\mathrm{reio},\ln(10^{10}A_\mathrm{s}),n_\mathrm{s},\Omega_\mathrm{g}\}$ of the $\Lambda$CDM and CGG ($\Lambda$CDM+$\Omega_\mathrm{g}$) models using Planck18 data ($TT,TE,EE$+low$E$+lensing likelihood). The $\Lambda$CDM model is equivalent to the CGG model with a fixed $\Omega_\mathrm{g}=0$.
    }
    \label{fig:Pl18CGG}
\end{figure}

\begin{figure}[tbp]
    \centerline{\includegraphics[width=0.8\textwidth]{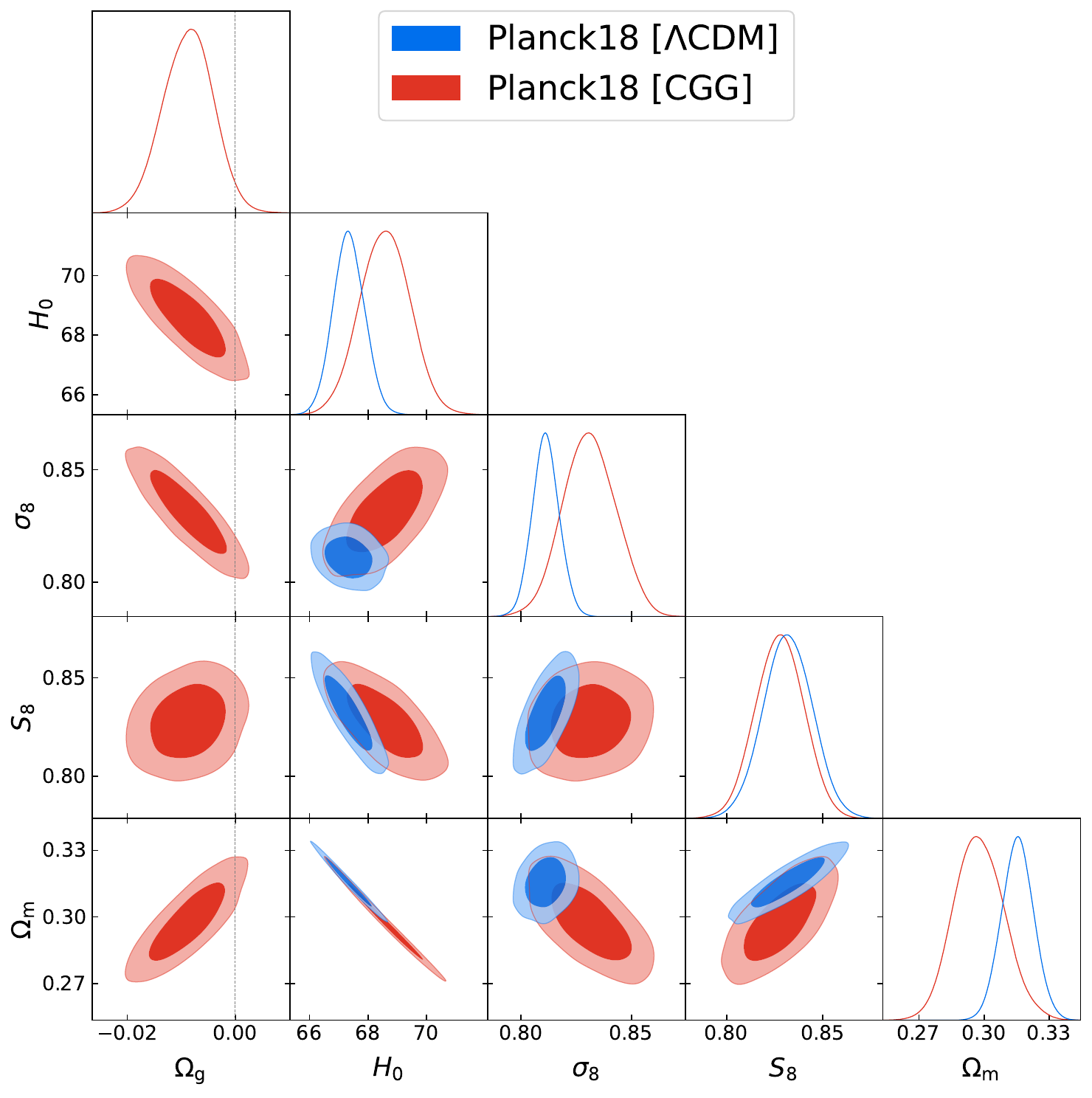}}
    \caption{Parameter constraints for the three derived cosmological parameters $\{\Omega_\mathrm{m},\sigma_8,S_8\}$, along with $H_0$ and $\Omega_\mathrm{g}$ of the $\Lambda$CDM and CGG ($\Lambda$CDM+$\Omega_\mathrm{g}$) models using the Planck18 data ($TT,TE,EE$+low$E$+lensing likelihood). In the $\Lambda$CDM model, $\Omega_\mathrm{g}=0$.
    }
    \label{fig:Pl18CGG_derived}
\end{figure}

Without a theoretically preferred $\Omega_\mathrm{g}$, we can constrain its value using cosmological observations of the CMB and large-scale structure (LSS). We shall first constrain our CGG model from \cref{eq:density-CGG}, which is a one parameter-extension to the 6-parameter $\Lambda$CDM model consisting of $\{\Omega_\mathrm{b}h^2,\Omega_\mathrm{c}h^2, H_0, \tau_\mathrm{reio}, A_s, n_s\}$,\footnote{We elect to use $H_0$ instead of $\theta_\ast$ as the third parameter, which is more appropriate when combining with other data sets such as SH0ES.} using the 2018 \textit{Planck} $TT,TE,EE$+low$E$+lensing likelihoods (hereafter referred to as Planck18)~\cite{18Plancklikelihood,18Planckparameter}. This can be compared to the 6-parameter $\Lambda$CDM fits. Our results were computed using the nested sampler \texttt{PolyChord}~\cite{15Handley_Polychord_stat,15Handley_polychord_cosmo} interfaced with \texttt{Cobaya}~\cite{2020Cobaya} and and a modified version of \texttt{CAMB}~\cite{CAMB} that implements our CGG model. The priors used in the the nested sampler are given in \cref{tab:prior}. To check the robustness of our constraints, we also perform a Markov chain Monte Carlo (MCMC) analysis through \texttt{Cobaya} for each model with broader priors compared to bounds given in \cref{tab:prior}, and the MCMC constraints agree with those obtained with nested sampling. We obtain parameter confidence intervals from the nested sampling results using \texttt{GetDist}~\cite{19GetDist}. The constraints of the main six parameters with and without $\Omega_\mathrm{g}$ are reported in \cref{tab:Constraints} and plotted in \cref{fig:Pl18CGG}, and the constraints for the three derived parameters of interest $\{\Omega_\mathrm{m},\sigma_8,S_8\}$ together with the constraints for $H_0$ and $\Omega_\mathrm{g}$ are shown in \cref{fig:Pl18CGG_derived}. 

\begin{table*}[tbp]
\centering
\begin{tabular}{ccccccc}
\hline
\hline
\noalign{\vskip 2pt}
 $\Omega_\mathrm{b} h^2$ & $\Omega_\mathrm{c} h^2$ & $H_0$ & $\tau$ & $\ln(10^{10}A_\mathrm{s})$ & $n_\mathrm{s}$ & $\Omega_\mathrm{g}$ \\
\hline
\noalign{\vskip 1pt}
[0.019,0.025] & [0.095,0.145] & [50,90] & [0.01,0.8] & [2.5,3.7] & [0.9,1.1] & [$-$0.06,0.06] \\
\hline
\end{tabular}
\caption{Lower and upper bounds of the uniform parameter priors used in the nested sampling presented in this work.}
\label{tab:prior}
\end{table*}

\begin{table}[tbp]
\setlength{\tabcolsep}{10pt}
\begin{center}
\begin{tabular}{l c c c c}
\hline
\hline
\noalign{\vskip 1pt}
Parameter & Planck18 [$\Lambda$CDM] & Planck18 [CGG] & Planck18 [CGG]\\
& & & best fit\\
\hline
\noalign{\vskip 2pt}
$\Omega_\mathrm{b} h^2$ & $0.02237\pm0.00014$ & $0.02248\pm0.00016$ & 0.0225\\
$\Omega_\mathrm{c} h^2$ & $0.1200\pm0.0012$ & $0.1168\pm0.0020$ & 0.1162\\
$H_0$ & $67.36\pm0.54$ & $68.58\pm0.86$ & 68.84\\
$\tau_\mathrm{reio}$ & $0.0542^{+0.0071}_{-0.0082}$ & $0.0499^{+0.0079}_{-0.0071}$ & 0.0507\\
$\ln(10^{10} A_\mathrm{s})$  & $3.044\pm0.015$ & $3.033\pm 0.016$ & 3.034\\
$n_\mathrm{s}$ & $0.9649\pm0.0042$ & $0.9690\pm0.0046$ & 0.971\\
$\Omega_\mathrm{g}$ & $0$ & $-0.0087\pm0.0046$ & $-$0.0098\\
\hline
$\Omega_\mathrm{m}$ & $0.3155\pm0.0074$ & $0.298\pm0.012$ & 0.294\\
$\sigma_8$ & $0.8112\pm0.0061$ & $0.831\pm0.012$ & 0.834\\
$S_8\equiv\sigma_8\sqrt{\Omega_\mathrm{m}/0.3}$ & $0.832\pm0.013$ & $0.828\pm0.012$ & 0.826\\
\hline
\end{tabular}
\end{center}
\caption{Parameter constraints for CGG versus $\Lambda$CDM. The first two columns show the mean and $1\,\sigma$ uncertainties for the cosmological parameters in $\Lambda$CDM and CGG models fit to the Planck18 data. The last column gives the CGG best-fit parameter values.}
\label{tab:Constraints}
\end{table}

\begin{figure}[tb]
\centerline{\includegraphics[width=0.93\hsize]{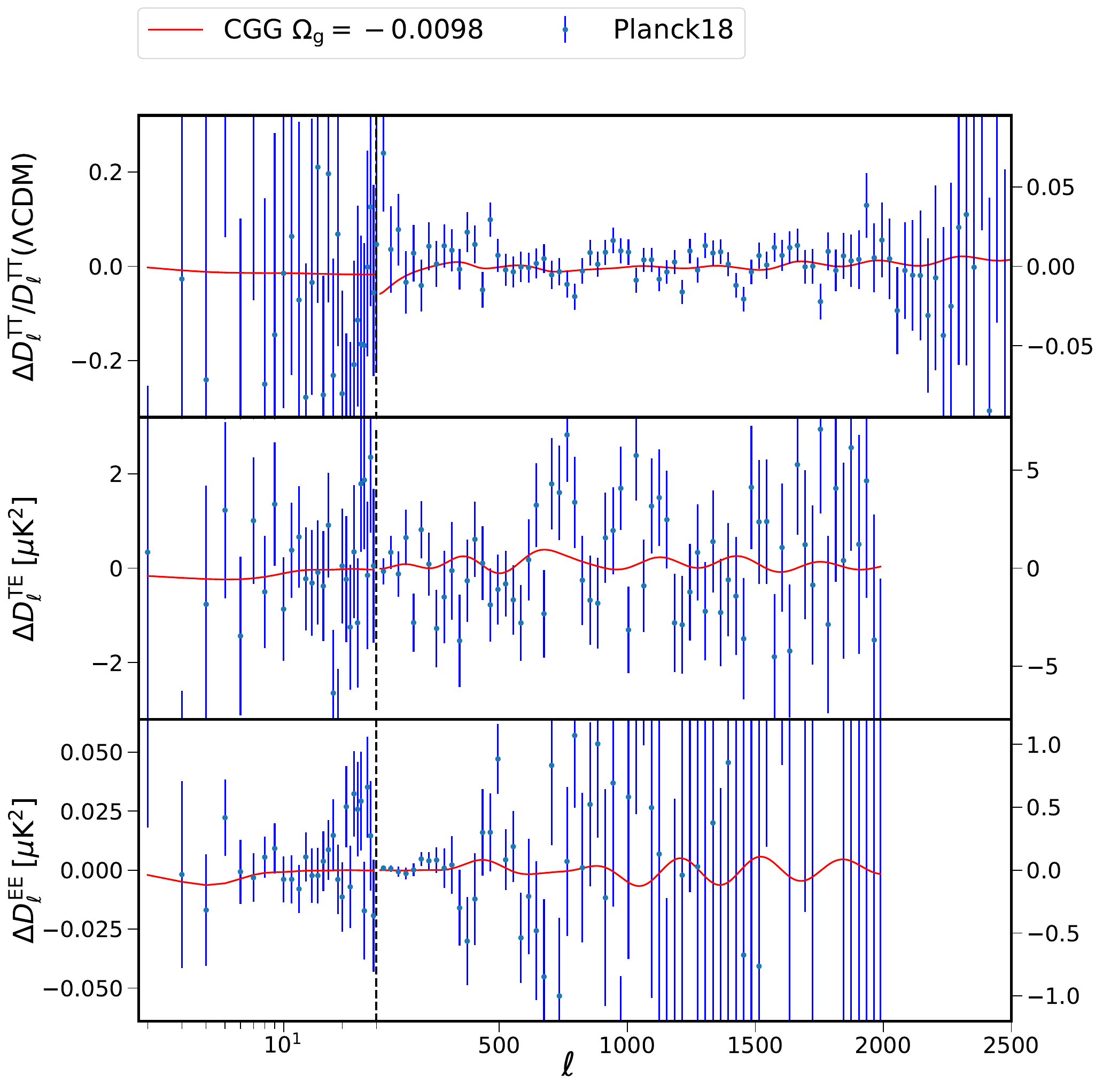}}
    \caption{Residuals of the CMB power spectra with respect to the best-fit $\Lambda$CDM model. Blue points show the residuals between the measured Planck18 power spectra (including error bars) and the best-fit $\Lambda$CDM model for Planck18 data, while the red line shows the residuals between the best-fit CGG model for the Planck18 data and the best-fit $\Lambda$CDM model. For the $TT$ power spectrum, we plot the percentage difference, while we show the absolute difference for the $TE$ and $EE$ power spectra. We set the $x$-axis to have a logarithmic scale for low multipoles $\ell\leq 30$ and a linear scale for high multipoles $\ell>30$; the corresponding $y$-axis labels are on the left and right sides, respectively.
}
\label{fig:Pl18bestfit}
\end{figure}

As seen in both \cref{tab:Constraints} and \cref{fig:Pl18CGG}, the 2018 \textit{Planck} data prefer $\Omega_\mathrm{g}$ in the negative region, with a value of order~1\,\%. The $\Lambda$CDM model, which corresponds to the CGG model with fixed $\Omega_\mathrm{g}=0$, is almost $2\sigma$ away from the mean $\Omega_\mathrm{g}$ value in CGG. The $\Omega_\mathrm{g}$ parameter is most degenerate with the matter density $\Omega_\mathrm{c}h^2$ and the Hubble parameter $H_0$, in such a way that  a negative $\Omega_\mathrm{g}$ allows a lower matter density at the present time and a higher Hubble expansion rate, while the constraints for the other four main cosmological parameters $\{\Omega_\mathrm{b}h^2,\tau_\mathrm{reio},\ln(10^{10}A_\mathrm{s}),n_\mathrm{s}\}$ remain essentially unchanged. The degeneracy between $\Omega_\mathrm{g}$ and $H_0$ can somewhat alleviate the Hubble tension, which we will discuss in \cref{sec:H0tension}.

To see why Planck18 data prefer a negative $\Omega_\mathrm{g}$ value, we examine the best-fit parameters in both $\Lambda$CDM and CGG, and compare the predicted power spectra of these best-fit results to the measured \textit{Planck} power spectra. As seen in \cref{tab:Best-Fit-Chi} and \cref{fig:Pl18bestfit}, the CGG model with negative $\Omega_\mathrm{g}$ lowers the low-$\ell$ part of the $TT$ and $EE$ power spectra, which provides a slightly better fit to the ``low-$\ell$ deficit''~\cite{Planckshifts,18Planckparameter} observed in the measured CMB power spectra. For the high-$\ell$ part of the $TT$ and $TE$ power spectra, the CGG model provides a slightly better fit to the apparent oscillations in the residuals between the measured Planck18 power spectra and the best-fit $\Lambda$CDM model (the first two panels of \cref{fig:Pl18bestfit}), which account for most of the overall $\chi^2$ improvement (the third row of \cref{tab:Best-Fit-Chi}). 

\begin{table}[tbp]
\setlength{\tabcolsep}{10pt}
\begin{center}
\begin{tabular}{l c c c}
    \hline
    \hline
    \noalign{\vskip 1pt}
    Data Set & $\Lambda$CDM & CGG & \phantom{$-$}$\Delta{\chi^2}$ \\
    \hline
    \noalign{\vskip 2pt}
    low-$\ell$ TT & 23.3 & 22.1 & $-1.2$\\
    low-$\ell$ EE  & 396.0 & 395.6 & $-0.4$\\
    high-$\ell$ TTTEEE & 2344.9 & 2342.1 & $-2.8$\\
    lensing &  8.9 & 9.7 & \phantom{$-$}0.8\\
    total & 2773.1 & 2769.5 & $-3.6$ \\
    \hline
\end{tabular}
\end{center}
    \caption{Values of $\chi^2$ for the best-fit $\Lambda$CDM and CGG models using the different \textit{Planck} 2018 data combinations, and the difference of $\chi^2$ for the two models ($\Delta \chi^2=\chi_\mathrm{CGG}-\chi_{\Lambda\mathrm{CDM}}$). The CGG model provides a better fit to the \textit{Planck} data overall.
    }
    \label{tab:Best-Fit-Chi}
\end{table}

If we restrict $\Omega_\mathrm{g}$ to the positive region, we find $\Omega_\mathrm{g}\,{<}\,0.002$ and $0.005$ for the $68\,\%$ and $95\,\%$ confidence intervals, which is consistent with the constraints on $\Omega_\mathrm{g}$ under similar EDE parametrizations~\cite{15PlanckDEMG,21Gomez-ValentEDEplateau}. Thus, we discover for the first time that the extremely tight constraint on $\Omega_\mathrm{g}$ for the EDE models in the literature is because \textit{Planck} CMB data---in fact---favour negative values.

\begin{figure}[tbp]
    \centerline{\includegraphics[width=0.75\hsize]{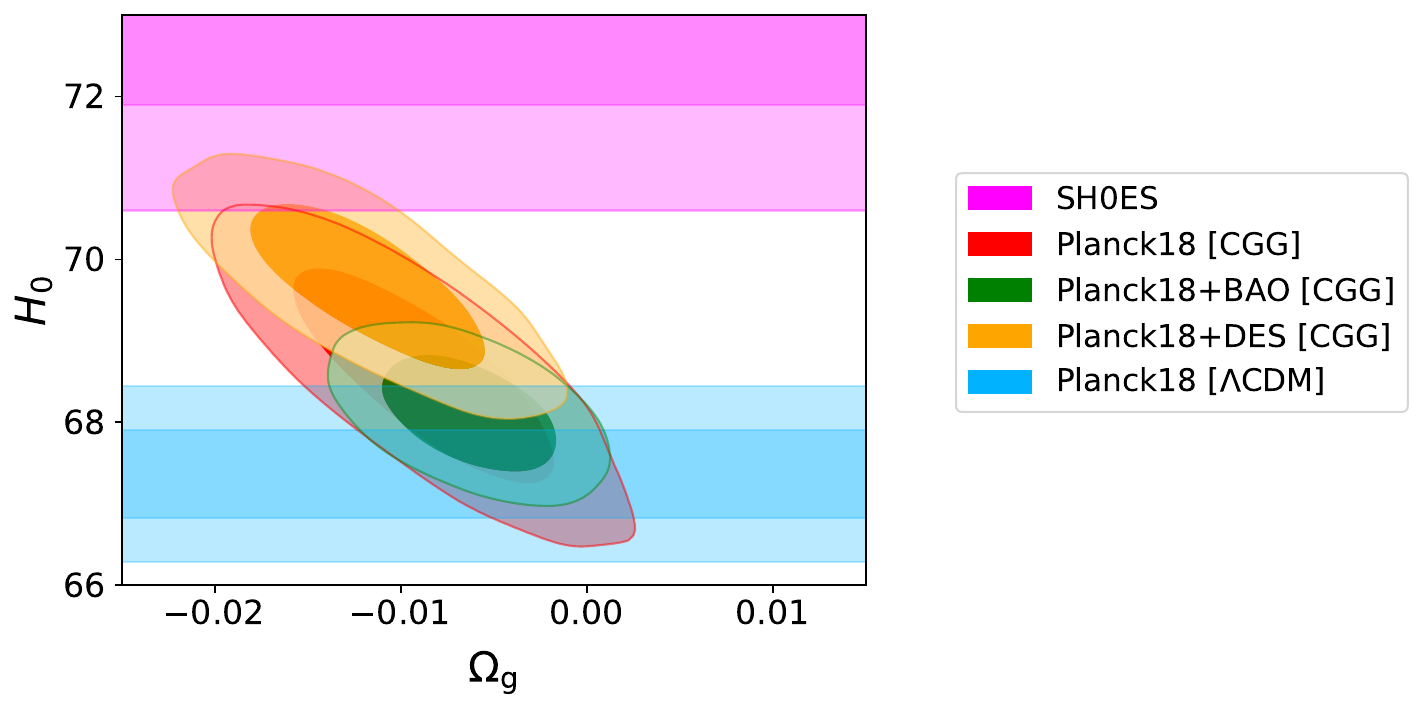}}
    \caption[Caption for Hubble]{Constraints for $H_0$ and $\Omega_\mathrm{g}$ in CGG ($\Lambda$CDM+$\Omega_\mathrm{g}$) using Planck18 ($TT,TE,EE$+low$E$+lensing), Planck18+BAO and Planck18+DES. The magenta band shows the distance-ladder measurement of $H_0$ from \textsc{SH0ES}~\cite{21SH0ES}, while the blue band shows the measurement of $H_0$ from the complete 2018 \textit{ Planck} data under the $\Lambda$CDM model.\protect\footnotemark \hspace{0.2cm}The CGG model significantly alleviates the Hubble tension between the Planck18 and SH0ES measurements. The constraint of $H_0$ from Planck18+DES under our CGG model is compatible with the SH0ES measurement.}
    \label{fig:CGG-H0}
\end{figure}

\subsection{Hubble tension}\label{sec:H0tension}

We now discuss the constraint on $H_0$ in our CGG model with Planck18 data in light of the Hubble tension. In recent years, the value of the Hubble constant $H_0$ inferred from probes of the early Universe from the CMB~\cite{18Planckparameter} has been in persistent disagreement with distance-ladder-derived estimates from the late Universe, a discrepancy that has now reached roughly the $5\,\sigma$ level~\cite{2019SH0ES,21SH0ES,22SH0ES}. For example, in~\cite{21SH0ES}, the Hubble constant is estimated to be $H_0=\SI{73.2\pm1.3}{\km\per\s\per\mega\parsec}$ (hereafter referred to as SH0ES) in contrast to the Planck18 value of $H_0=\SI{67.36\pm0.54}{\km\per\s\per\mega\parsec}$.

With our CGG model under Planck18, $H_0$ is determined to be $\SI{68.58\pm0.86}{\km\per\s\per\mega\parsec}$, which is higher than the $\Lambda$CDM value, thus somewhat alleviating the Hubble tension. The posterior average of $\Omega_\mathrm{g}$ is $-0.0087\pm0.0046$. The joint constraints for $H_0$ and $\Omega_\mathrm{g}$ are plotted in \cref{fig:CGG-H0}, which shows the strong degeneracy between the two parameters. Combining the Planck18 data with baryonic acoustic oscillation (BAO) measurements~\cite{11sixdf_2011_bao,15sdss_dr7_mgs,17BAODR12}, we obtain $H_0=\SI{68.11\pm0.46}{\km\per\s\per\mega\parsec}$ and $\Omega_\mathrm{g}=-0.0063\pm0.0031$ for our CGG model, which is more consistent with the $\Lambda$CDM results compared to the CGG fits using the Planck18 data only. Combining the Planck18 data with galaxy clustering and weak gravitational lensing measurements from the Dark Energy Survey~(DES)~\cite{18DES_Y1}, we obtain $H_0=\SI{69.69\pm0.66}{\km\per\s\per\mega\parsec}$ and $\Omega_\mathrm{g}=-0.0118\pm0.0042$, which is slightly higher than the $H_0$ measurements from our CGG model using Planck18 data only. These constraints are summarised in \cref{fig:CGG-H0}. {Adding the Pantheon Supernovae (SNe) likelihood~\cite{18Sne} to the Planck18 data only very slightly shrinks the constraints on $\Omega_{\rm g}$ and $H_0$, with $\Omega_\mathrm{g}=-0.0083\pm0.0041$ and $H_0=\SI{68.55\pm0.74}{\km\per\s\per\mega\parsec}$, leaving the corresponding contours from \cref{fig:Pl18CGG} essentially indistinguishable.

Our CGG model with only one additional parameter eases the Hubble tension (from 4.1$\sigma$ to 3.0$\sigma$) when we only consider the 2018 {\it Planck} measurements of the CMB. Allowing $\Omega_\mathrm{g}$ to be negative helps to reconcile the Hubble tension, since a lower $\Omega_\mathrm{g}$ corresponds to a higher $H_0$ value. Adding DES Y1 data further increases the preferred $H_0$, to within $2.4\,\sigma$ of the local SH0ES measurement. However, adding BAO data to the Planck18 measurements tightens the constraint of $H_0$, while lowering the mean value to $3.7\,\sigma$ away from the value in Ref.~\cite{22SH0ES}. The summary is that although the CGG model somewhat alleviates the $H_0$ tension, the situation is complicated and depends on the choice of data set.  Moreover, we need to consider the statistical effect of adding an additional parametric degree of freedom to the fits, which we will discuss in \cref{sec:model-comparison}.

\footnotetext{Strictly speaking, using a horizontal band (the blue band) to show the $H_0$ measurement from the full 2018 \textit{Planck} data under the $\Lambda$CDM model is not correct, since $\Omega_\mathrm{g}$ is fixed at zero in this case. We use only the horizontal band here for illustrative purposes.}

\begin{figure}  \centerline{\includegraphics[width=0.75\hsize]{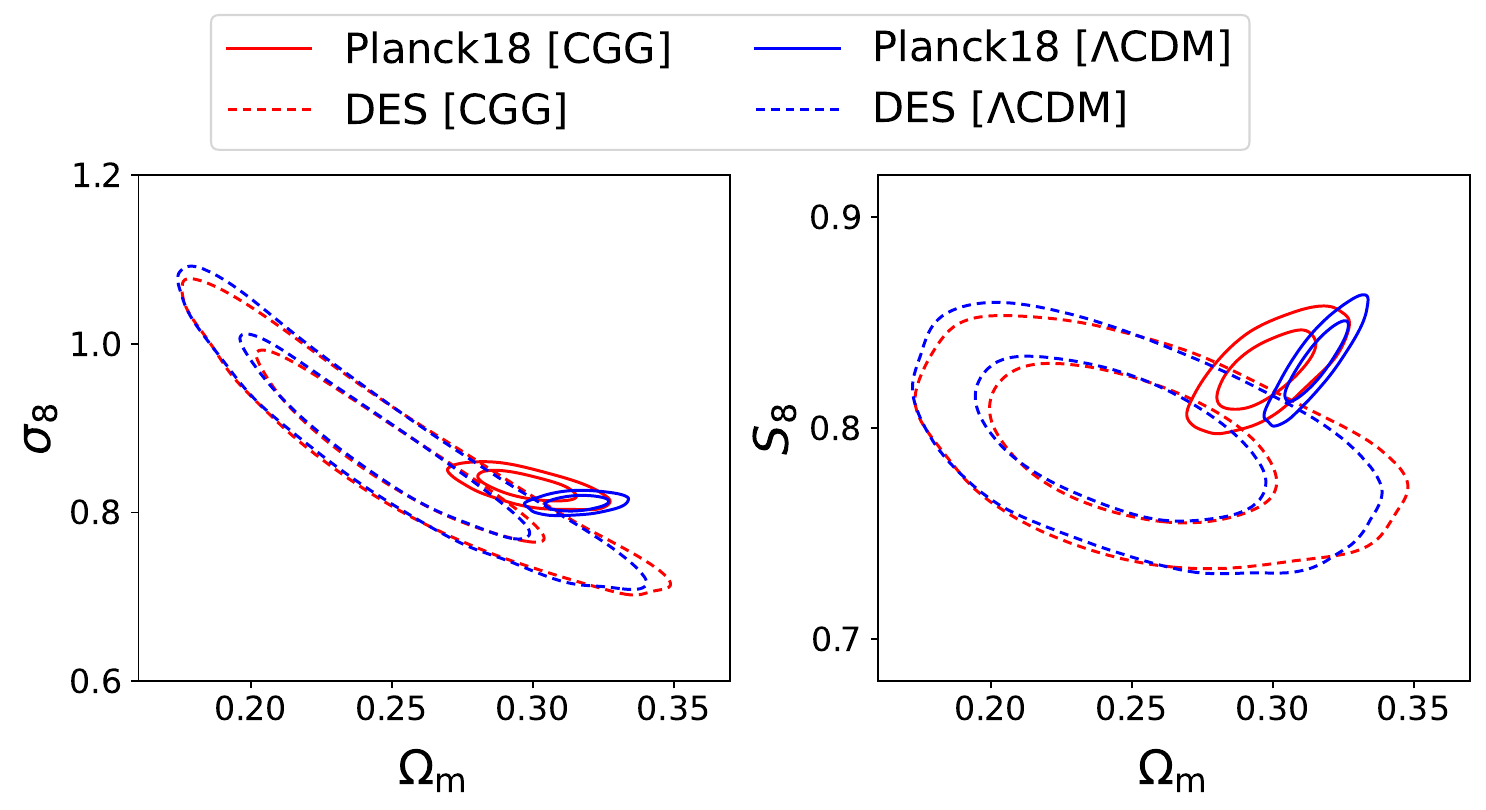}}
    \caption{
        Constraints on $S_8$, $\sigma_8$ and $\Omega_\mathrm{m}$ in the $\Lambda$CDM and CGG models using Planck18 ($TT,TE,EE$+low$E$+lensing) only and DES data only. The solid and dashed line styles correspond to Planck18 and DES data, respectively, while the red and blue colors correspond to the CGG and $\Lambda$CDM models, respectively. We see that the clustering tension in terms of the 2D $S_8\text{-}\Omega_\mathrm{m}$ plane is alleviated by the CGG model due to a decrease in the $\Omega_\mathrm{m}$ constraint under the Planck18 data.
    }
    \label{fig:CGG-S8}
\end{figure}
\subsection{Clustering tension}
\label{sec:S8tension}

As illustrated in \cref{fig:Pl18CGG_derived}, $\Omega_\mathrm{g}$ is also correlated in the fits with $\sigma_8$ and $\Omega_\mathrm{m}$.  However, the negative $\Omega_\mathrm{g}$ value increases the fitted $\sigma_8$ value compared to the $\Lambda$CDM results, which does not alleviate the $\sigma_8$ tension. To obtain a lower $\sigma_8$ value, we would need a positive $\Omega_\mathrm{g}$, which is disfavoured by the Planck18 data. Adding massive neutrinos as a free parameter to our CGG model would be one way to help with the $\sigma_8$ tension, since increasing the mass of neutrinos will lead to a lower $\sigma_8$ constraint based on the \textit{Planck} data. Turning to the $S_8$ tension, our model does not shift the $S_8$ constraint, since $\sigma_8$ and $\Omega_\mathrm{g}$ receive opposite shifts that essentially cancel in the $\sigma_8\sqrt{\Omega_\mathrm{m}/0.3}$ combination. Therefore, the CGG model neither alleviates nor worsens the $S_8$ tension, which is similar to other EDE models~\cite{21MurgiaEDE}. Note that $S_8$ is anti-correlated with $H_0$, so a lower $S_8$ value corresponds to a higher $H_0$ value, as seen in \cref{fig:Pl18CGG_derived}. 

Considering the 2-dimensional (2D) representation of the clustering tension, we plot the constraints on $\Omega_\mathrm{m}$ and $S_8$ using Planck18 data and DES data in \cref{fig:CGG-S8}. DES data alone have little constraining power on $\Omega_\mathrm{g}$, and the constraints of $\Omega_\mathrm{m}$ and $S_8$ for DES under the CGG model are similar to those under $\Lambda$CDM. However, the CGG model noticeably lowers the best-fit for $\Omega_\mathrm{m}$ under the Planck18 data by allowing negative $\Omega_\mathrm{g}$ values (\cref{fig:Pl18CGG_derived}). As a result, the $S_8$--$\Omega_\mathrm{m}$ confidence region of DES and that of Planck18 have a larger area of overlap under the CGG model compared to the $\Lambda$CDM case, as seen in the right panel of \cref{fig:CGG-S8}. Therefore, the clustering tension is in fact slightly alleviated by the CGG model when we view the tension in the 2D plane formed by $S_8$ and $\Omega_\mathrm{m}$ instead of only considering the single parameter $S_8$. The situation for the $\sigma_8$--$\Omega_\mathrm{m}$ 2D plane is similar (left panel in \cref{fig:CGG-S8}).

\subsection{Model comparison}
\label{sec:model-comparison}
To properly account for the increased volume of the parameter space under the CGG model over $\Lambda$CDM, we perform a Bayesian model comparison by calculating the difference of Bayesian evidence $\Delta \ln \mathcal{Z}=\ln\mathcal{Z}_\mathrm{CGG}-\ln \mathcal{Z}_{\Lambda\mathrm{CDM}}$, which naturally incorporates the so-called Occam’s razor that penalises models for unnecessary complexity. For Bayesian model comparison, we follow the procedure outlined in Ref.~\cite{21HergtModelComp} using the nested sampler \texttt{PolyChord}~\cite{15Handley_Polychord_stat,15Handley_polychord_cosmo} to explore the parameter space (the priors of the cosmological parameters are specified in \cref{tab:prior}), and using \texttt{anesthetic}~\cite{19Anesthetic} for the computation of Bayesian evidence, Kullback--Leibler divergence, and further statistics. 

\begin{figure}
    \centering
    \begin{subfigure}{0.32\textwidth}
        \centering
        \includegraphics[width=\textwidth]{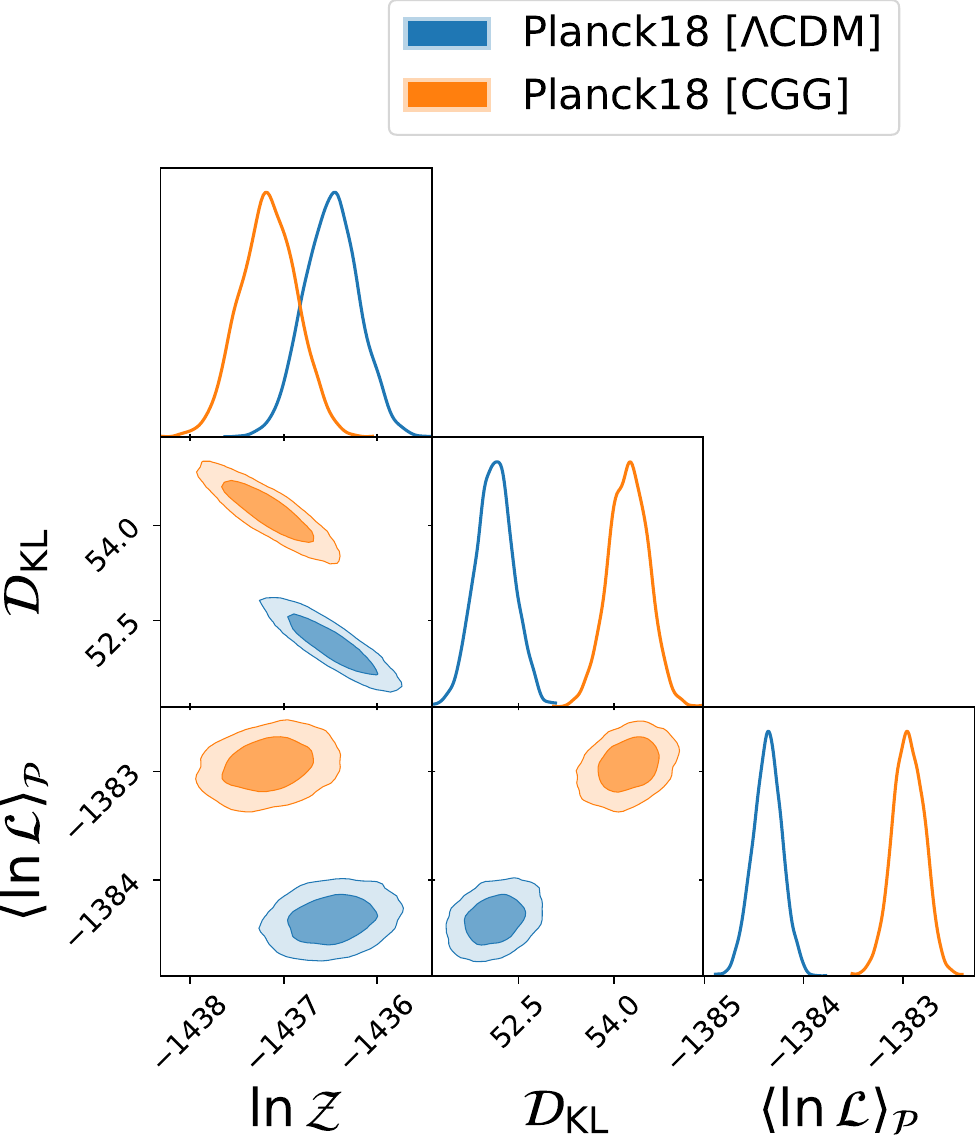}
        \caption{}
        \label{fig:plBM}
    \end{subfigure}
    \hfill
    \begin{subfigure}{0.32\textwidth}
        \centering
        \includegraphics[width=\textwidth]{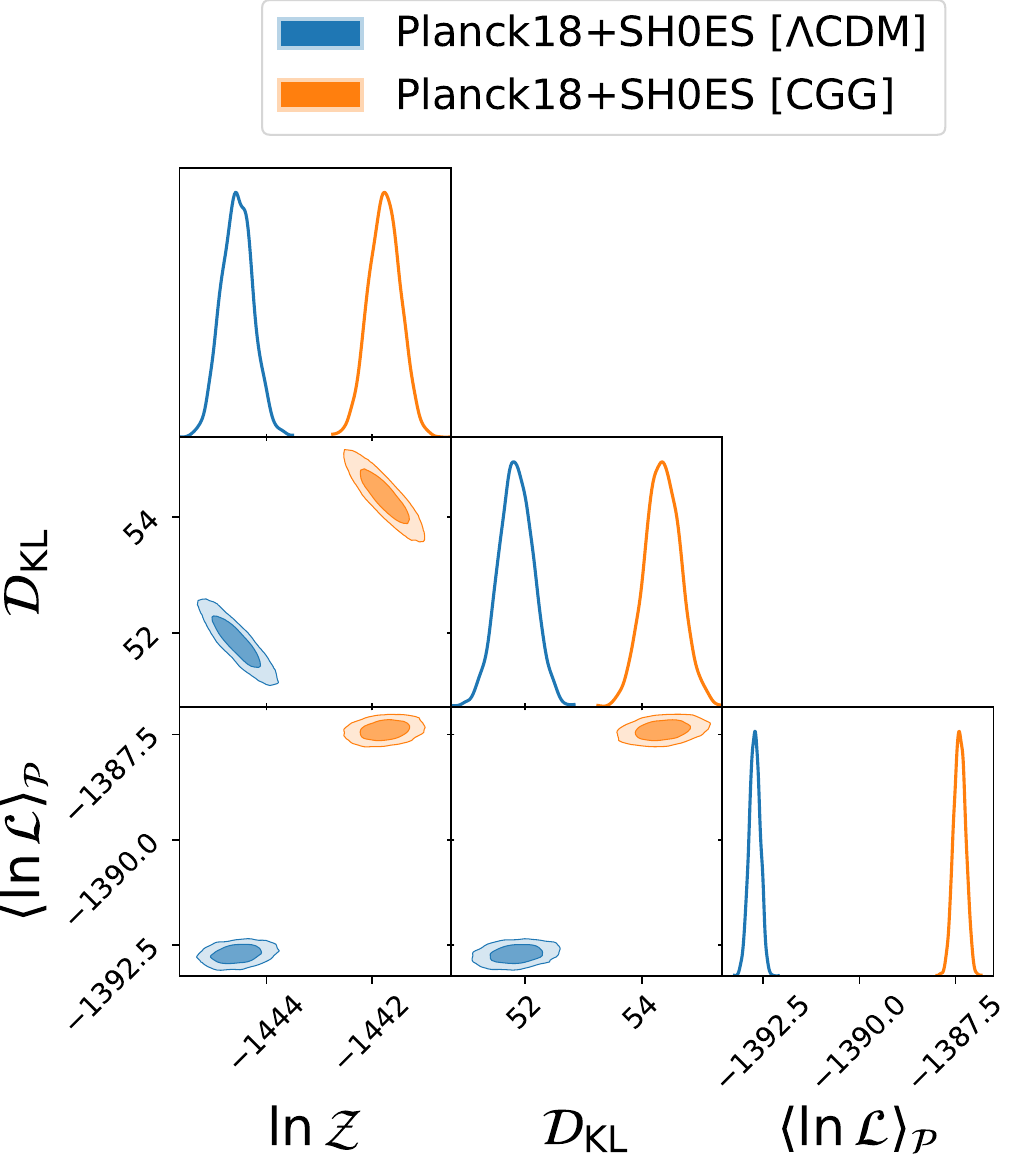}
        \caption{}
        \label{fig:plH0BM}
    \end{subfigure}
\hfill
    \begin{subfigure}{0.32\textwidth}
        \centering
        \includegraphics[width=\textwidth]{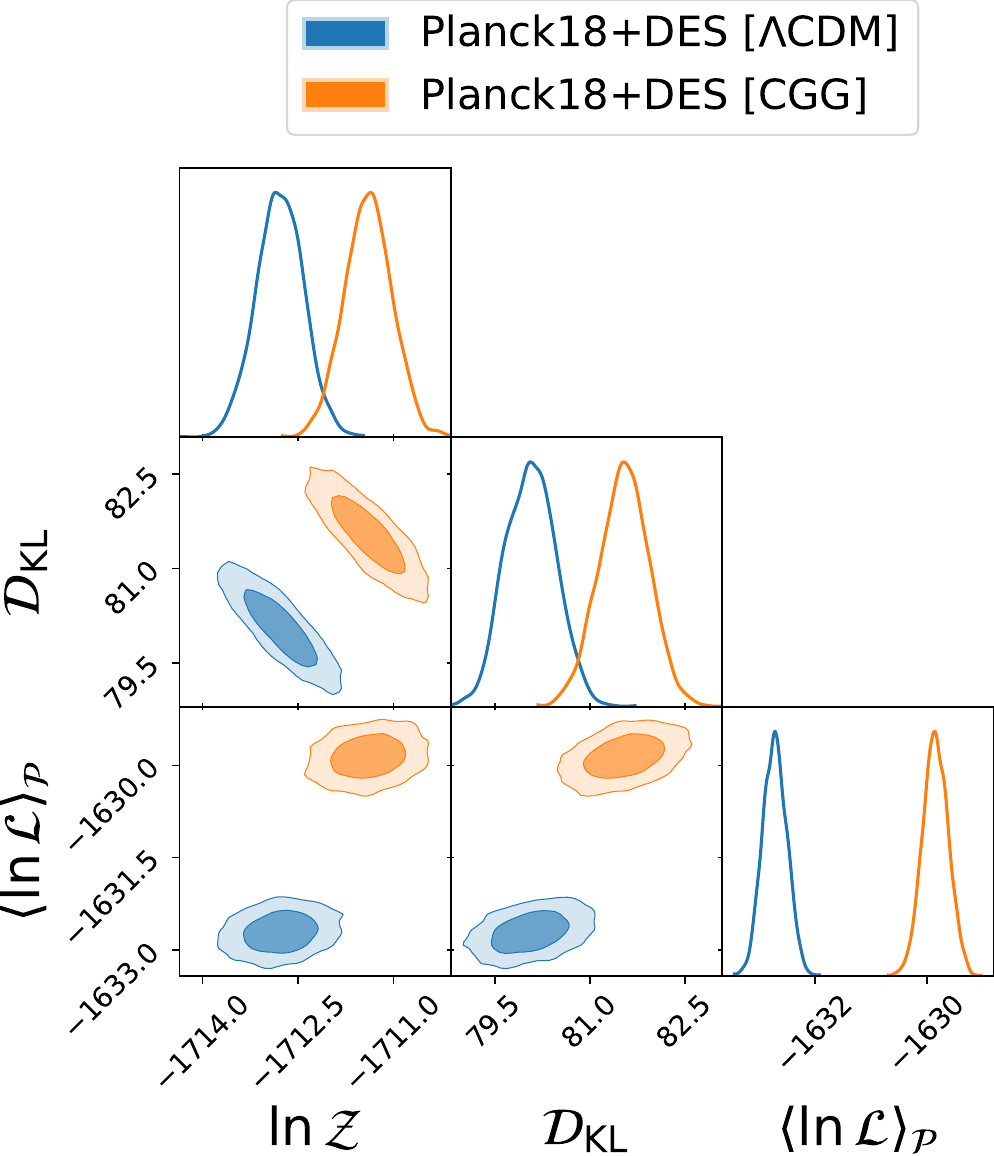}
        \caption{}
        \label{fig:plDESBM}
    \end{subfigure}
    \caption{Bayesian model comparison between the $\Lambda$CDM model with $\Omega_\mathrm{g}=0$ in blue and the CGG model with variable $\Omega_\mathrm{g}$ in orange, under different likelihoods. We plot the log-evidence $\ln\mZ$, Kullback–Leibler divergence $\mD_\mathrm{KL}$ (a measurement of the Occam penalty caused by increasing model complexity) and posterior average of the log-likelihood $\langle\ln\mL\rangle_\mP=\ln\mZ+\mD_\mathrm{KL}$ (indicating the fit of the model). The distributions here represent the sampling uncertainty from the nested sampling process. In the limit of infinite sampling precision these distributions would become point statistics.}
\end{figure}

As seen in \cref{fig:plBM}, although the CGG model provides a better fit to the Planck18 data, the improvement in the fit $\langle\ln\mL\rangle_\mP$ is not sufficient to compensate for the increase in complexity measured by the Kullback--Leibler divergence $\mD_\mathrm{KL}$. The two models are on-par with each other under the Planck18 likelihood in terms of Bayesian evidence~$\ln\mZ$. However, if we perform the Bayesian model comparison under both the Planck18 and SH0ES likelihoods (\cref{fig:plH0BM}), we see that the log-evidence $\ln\mZ$ under the CGG model is higher than that under $\Lambda$CDM, with a significantly better fit $\Delta\av[\mP]{\ln\mL}\approx5$ overcoming the penalty from the higher complexity of having an additional parameter. In this case, we have $\Delta\ln\mZ\gtrsim2$, suggesting that the CGG model is clearly favoured (although not conclusively), over the $\Lambda$CDM model under the Planck18+SH0ES likelihood. This reflects how the $\Omega_\mathrm{g}$ extensions helps alleviate the Hubble tension between Planck18 and SH0ES.

As for the model performance under the DES data only, the statistics considered here have consistent values, up to sampling uncertainties, for the CGG and the $\Lambda$CDM models. Under both the Planck18 and DES likelihoods, we see in \cref{fig:plDESBM} that the CGG model is slightly preferred over $\Lambda$CDM by $\Delta\ln\mZ\approx1.5$ (although the sampling uncertainty for the $\ln\mZ$ estimate here is about 0.5), which indicates the slight alleviation of the clustering tension based on model comparison. Using suspiciousness $\ln\mS\equiv\av[\mP]{\ln\mL_{AB}} - \av[\mP]{\ln\mL_A} - \av[\mP]{\ln\mL_B}$ \cite{19Handley,21Handley_ACTSPT}, which can quantify the global tension between two dataset $A$ and $B$ over the entire parameter space, the tension between Planck18 and DES reduces from 2.2 $\sigma$ ($\Lambda$CDM) to 2 $\sigma$ (CGG), which reflects the slight alleviation of the clustering tension by adding $\Omega_\mathrm{g}$.

\section{Impacts of \textit{Planck} PR4 data}\label{sec:PR4}

Since this work was completed, a new analysis of the Planck Public Release~4 (PR4) data set \cite{20Npipe} has been published, using the \texttt{LoLLiPoP} (low-$\ell$) and \texttt{HiLLiPoP}  (high-$\ell$) likelihoods \cite{Tristram2023}. This new likelihood analysis of \textit{Planck} data yields tighter constraints, as well as slightly shifting results for some previous claims of parameter anomalies (e.g.\ the lensing consistency parameter $A_{\rm L}$) \cite{Tristram2023}. For this reason, it is useful to check how much the constraints on the CGG model change if we use the PR4 data rather than the Planck18 data (also known as PR3).

We have derived cosmological constraints on the CGG model using the \texttt{HiLLiPoP} (high-$\ell$ TTTEEE), \texttt{LoLLiPoP} (low-$\ell$ EE), and CMB lensing likelihoods \cite{22Carron_PR4lensing} based on \textit{Planck} PR4. The only part of the PR4 analysis still relying on the Planck18 (PR3) data is the low-$\ell$ temperature likelihood. Under the PR4 analysis, all parameter constraints of the CGG model are slightly tighter compared to the PR3 analysis. In particular, we find that $\Omega_{\rm g}=-0.0054\pm 0.0042$ and $H_0=\SI{68.20\pm0.71}{\km\per\s\per\mega\parsec}$. The $\Lambda$CDM model, which corresponds to the CGG model with fixed $\Omega_\mathrm{g}=0$, is now $1.3\,\sigma$ away from the mean $\Omega_\mathrm{g}$ value in CGG under the PR4 analysis, compared to the $1.9\,\sigma$ preference for negative $\Omega_\mathrm{g}$ values under the PR3 analysis. It is interesting to notice that the shift of $\Omega_\mathrm{g}$ (from $1.9$ to $1.3\,\sigma$) is significantly less than the shifts of $A_{\rm L}$ and $\Omega_K$ from the PR3 to PR4 analysis observed in Ref.~\cite{Tristram2023}. The PR4 likelihood still prefers negative $\Omega_{\rm g}$ values, albeit to a slightly less extent compared to the PR3 results.

For the clustering parameters, \textit{Planck} PR4 is known to relieve the clustering tension compared to PR3~\cite{Tristram2023} already for the $\Lambda$CDM model. This is still the case for the CGG model, alhough to a smaller degree, since the tension is already considerably decreased by switching from $\Lambda$CDM to CGG. For the clustering parameters under the CGG model, we find $\sigma_8=0.823\pm 0.011$ and $S_8=0.825\pm 0.011$.

\section{Forecasts for future CMB and BAO measurements}\label{sec:forecast}

\begin{table}[tb]
    \setlength{\tabcolsep}{10pt}
    \begin{center}
    \begin{tabular}{l c c c}
        \hline
        \hline
        \noalign{\vskip 1pt}
        Parameter & Current CMB & Ideal CMB & Ideal CMB+Euclid BAO\\
        \hline
        \noalign{\vskip 2pt}
        $\sigma(\Omega_\mathrm{g})$& $0.0046$ & $0.0011$ & $0.0008$ \\
        \hline
    \end{tabular}
    \end{center}
    \caption{Estimated uncertainties for $\Omega_\mathrm{g}$ from a 7-parameter Fisher-matrix calculation using the $TT$, $TE$ and $EE$ power spectra, assuming that we are in the cosmic-variance limit.  In this calculation, we go to a maximum multipole of $\ell_\mathrm{max}=3000$ for the $TT$ power spectrum and $\ell_\mathrm{max}=6000$ for the $TE$ and $EE$ power spectra, together with the lensing reconstruction spectra $C_{\ell}^{\phi\phi}$ for $\ell$ up to 1000. To examine the constraining power of future LSS experiments, we also add a \textit{Euclid}-like measurement of the BAO scale to the ideal CMB case. More details about the forecast are given in Appendix~A of Ref.~\cite{21WenT0}. The fiducial $\Lambda$CDM and $\Omega_\mathrm{g}$ models are taken from the best-fit results in \cref{sec:posterior-results}.}
    \label{tab:Omgea_g-forecast}
\end{table}

As discussed in \cref{sec:posterior-results,sec:PR4}, the current cosmological data favour a negative $\Omega_\mathrm{g}$ value. To distinguish whether $\Omega_\mathrm{g}$ is truly negative or consistent with 0, we will need better data.  So we now consider the constraining power of future cosmological measurements on $\Omega_\mathrm{g}$. To give a simplistic assessment of this idea, we combine the Fisher forecasts from an ideal cosmic variance limited (CVL) CMB experiment and the measurements of the BAO scale in a \textit{Euclid}-like survey~\cite{Euclid2011}. The details of the Fisher forecast method are described in Appendix~A of Ref.~\cite{21WenT0}. For the CVL CMB forecast, we make use of the $TT$, $TE$ and $EE$ power spectra ($\ell$ up to $3000$ for $TT$ and $6000$ for $TE,EE$) along with the lensing reconstruction spectra $C_{\ell}^{\phi\phi}$ for $\ell$ up to 1000. As shown in \cref{tab:Omgea_g-forecast}, the CVL CMB alone can achieve a $1\,\sigma$ uncertainty of around~$10^{-3}$, while adding \textit{Euclid}-like BAO data can improve the constraint to below~$10^{-3}$, which is approximately a four fold reduction of uncertainty compared to the current constraints. With such small uncertainties, we should be able to distinguish whether $\Omega_\mathrm{g}$ is truly negative or consistent with $\Lambda$CDM in the upcoming stage IV CMB~\cite{16CMBS4} and LSS surveys~\cite{Euclid2011,2016DESIintro}. Since our model mostly modifies the equation of state for dark energy~$w_\mathrm{DE}$ at high redshift, we expect that future LSS surveys \cite{19PUMA,19MegaMapper} focusing on $z>2$ will be particularly promising for constraining the cosmic glitch parameter~$\Omega_\mathrm{g}$.

\begin{figure}
    \centering\includegraphics{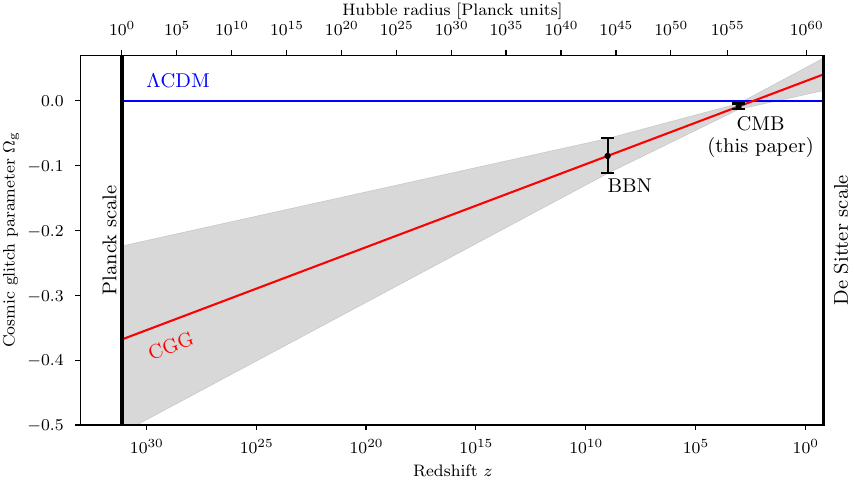}
    \caption{Current measurements of a non-constant cosmic glitch parameter $\Omega_\mathrm{g}$, as a function of Hubble radius at the time of the measurement. The lines and grey region show the linear extrapolation ($\pm1\,\sigma$) out to Planck and de Sitter (or dark energy) scales, suggesting that the glitch may vanish as our Universe approaches the de Sitter phase, while it is ${\cal O}(1)$ at the Big Bang, where curvature approaches the Planck scale.}
    \label{fig:running}
\end{figure}

\section{The possibility of non-constant \texorpdfstring{$\boldsymbol{\Omega}_\mathrm{g}$}{Omegag}}
\label{sec:discussion}
In addition to the $H_0$ and $S_8$ tensions, there may be differences in the early Universe. We noted in the introduction that an independent piece of evidence for a negative $\Omega_\mathrm{g}$ comes from measurements of helium abundance in extremely low metallicity galaxies, and its incompatibility with nucleosynthesis in the standard model~\cite{22EMPRESSVIIICuscuton}.  However, the value of $\Omega_\mathrm{g}=-0.085\pm0.027$ required to reconcile the measurements is significantly smaller than the constraints found here coming from fits to the \textit{Planck} 2018 data.\footnote{Incidentally, we checked that having a different $\Omega_\mathrm{g}$ at the Big Bang nucleosynthesis epoch, which would change the primordial helium abundance $Y_\mathrm{p}$, has a negligible effect on the parameter constraints presented in \cref{tab:Constraints}.} Given that these values affect comic dynamics in vastly different eras, one may speculate about a logarithmic running of the cosmic glitch with scale, similar to other fundamental dimensionless constants in renormalizable theories. Figure~\ref{fig:running} shows the values of these $\Omega_\mathrm{g}$ measurements, along with the possible linear extrapolations expected from a renormalisation group flow. This suggests a possible scenario where the glitch vanishes on the scale of the observed cosmological constant (or de~Sitter radius) today, where we recover near-exact de~Sitter symmetry. In contrast, at the Big Bang (or Planck scale), the glitch parameter is ${\cal O}(1)$, pointing to a significant violation of general covariance in a quantum theory of gravity. This may further hint at a genuine quantum gravity solution to the cosmological horizon problem, which is traditionally solved using an inflationary paradigm (e.g.~\cite{afshordi2016critical}).

\section{Conclusions}\label{sec:conclusion}

In this paper, we have examined the theoretical and observational cases for a {\it cosmic glitch in gravity}, i.e.\ a model in which gravity is different for super-horizon and sub-horizon scales, as a minimal modification of Einstein's general relativity without introducing any new scale or degree of freedom.  The best-fit CGG model prefers a higher $\sigma_8$ and lower $\Omega_\mathrm{m}$ than $\Lambda$CDM, but the value of $S_8$ has hardly changed.  On the other hand, we find that the \textit{ Planck} 2018 data favour a negative cosmic glitch parameter $\Omega_\mathrm{g}$, a parameter region that has not been explored before. The significance of the evidence for this glitch ranges from 1.9$\,\sigma$ to 2.8$\,\sigma$, depending on the additional large-scale structure data used in the analysis, while it decreases to 1.3$\,\sigma$ when using only CMB data and replacing the \textit{Planck} 2018 (PR3) likelihoods with the newer likelihoods based on \textit{Planck} PR4. We provide a summary of the constraints on $\Omega_\mathrm{g}$ in \cref{fig:omegag}.

\begin{figure}
    \centerline{\includegraphics[width=0.7\hsize]{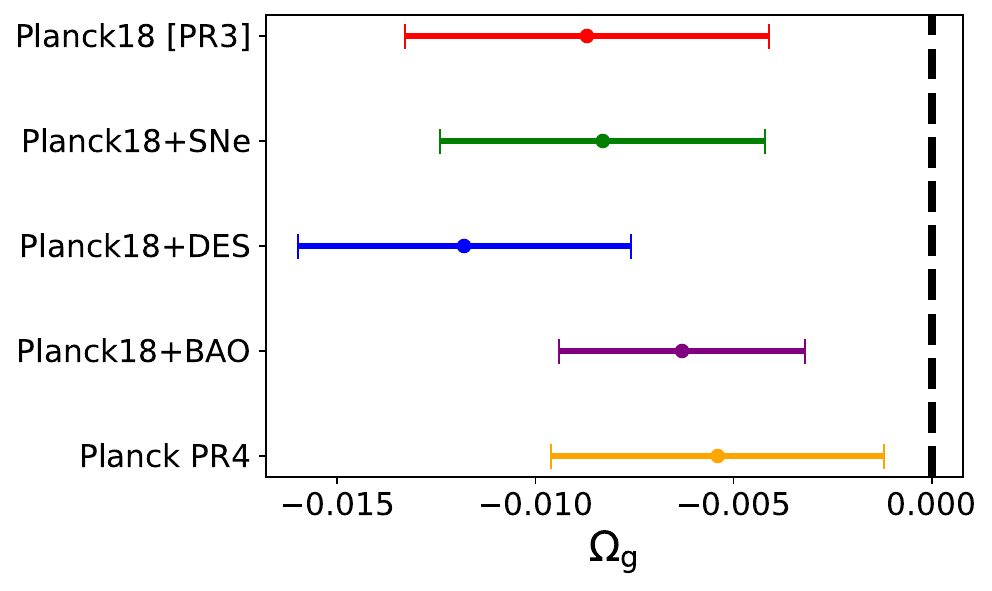}}
    \caption{Compilation of constraints on the cosmic-glitch parameter $\Omega_\mathrm{g}$ uisng different data combinations.}
    \label{fig:omegag}
\end{figure}

By analysing the parameter constraints and performing a Bayesian model comparison, we see that our CGG model somewhat alleviates both the Hubble parameter and the clustering tensions when using the \textit{Planck} 2018 data, while the $H_0$ constraint using the \textit{ Planck} 2018 and DES~Y1 data is compatible with the local SH0ES measurement. In contrast, including the observed BAO scale spoils this agreement on $H_0$. However, it is possible that current BAO scale measurements may be biased or their uncertainties might be underestimated for this class of non-$\Lambda$CDM cosmology~\cite{19Anselmi_BAO,22Anselmi_BAO}, something that requires further calibration in mock observations.  Nevertheless, this effectively negative early dark energy component, realised through the CGG, deserves more study.  Future CMB and large-scale structure data (such as DESI or \textit{Euclid}) will inevitably tighten these bounds and shed light on whether a cosmic glitch in gravity is responsible for some of our current cosmic tensions.

\acknowledgments
This research was supported by the Natural Sciences and Engineering Research Council of Canada. 
NA is further supported by the Perimeter Institute for Theoretical Physics.
Research at Perimeter Institute is supported in part by the Government of Canada through the Department of Innovation, Science and Economic Development Canada and by the Province of Ontario through the Ministry of Colleges and Universities. 
LTH was supported by a Killam Postdoctoral Fellowship and a CITA National Fellowship.
Computing resources were provided by the Digitial Research Alliance of Canada/Calcul Canada (\url{alliancecan.ca}). 
Parts of this paper are based on observations obtained with \textit{Planck} (\url{www.esa.int/Planck}), an ESA science mission with instruments and contributions directly funded by ESA Member States, NASA and Canada. This paper made use of the codes
\texttt{CAMB} (\url{camb.readthedocs.io/en/latest/}) and \texttt{Cobaya} (\url{cobaya.readthedocs.io/en/latest/}).

\bibliographystyle{JHEP}
\bibliography{refs}

\end{document}